\documentclass[bibyear]{aa}  
\usepackage[dvipsnames]{xcolor}

\usepackage{txfonts}
\usepackage{natbib}
\usepackage{xcolor}
\usepackage{url}
\usepackage{multicol}
\usepackage{multirow}
\usepackage[colorlinks=true,allcolors=blue]{hyperref}
\usepackage[utf8]{inputenc}
\usepackage{amsmath}
\usepackage{nccmath}
\usepackage{amsfonts}
\usepackage{amssymb}
\usepackage{graphicx}%pour pouvoir mettre des images avec includegraphics
\usepackage{comment}
\usepackage{caption}
\usepackage{subcaption}
\usepackage{soul}
\usepackage[normalem]{ulem}
\usepackage{gensymb}%to have the \degree command

\newcommand{\lenstro}{\texttt{lenstronomy}}

\begin{document}

\title{The impact of mass map truncation on strong lensing simulations}

\author{Lyne Van de Vyvere\inst{1}\fnmsep \thanks{\email{lyne.vandevyvere@uliege.be}}
          \and
          Dominique Sluse\inst{1}
          \and
          Sampath Mukherjee \inst{1}
          \and
          Dandan Xu \inst{2}
          \and
          Simon Birrer \inst{3}
          }

\institute{STAR Institute, Quartier Agora - All\'ee du six Ao\^ut, 19c B-4000 Li\`ege, Belgium \and Department of Astronomy, Tsinghua University, Beijing, 100084, China \and Kavli Institute for Particle Astrophysics and Cosmology and Department of Physics, Stanford University, Stanford, CA 94305, USA}

   \date{Received 16--07--2020; accepted 11--10--2020 }

  \abstract
   {Strong gravitational lensing is a powerful tool to measure cosmological parameters and to study galaxy evolution mechanisms. However, quantitative strong lensing studies often require mock observations. To capture the full complexity of galaxies, the lensing galaxy is often drawn from high resolution, dark matter only or hydro-dynamical simulations. These have their own limitations, but the way we use them to emulate mock lensed systems may also introduce significant artefacts. In this work we identify and explore the specific impact of mass truncation on simulations of strong lenses by applying different truncation schemes to a fiducial density profile with conformal isodensity contours. Our main finding is that improper mass truncation can introduce undesired artificial shear. The amplitude of the spurious shear depends on the shape and size of the truncation area as well as on the slope and ellipticity of the lens density profile. Due to this effect, the value of $H_0$ or the shear amplitude inferred by modelling those systems may be biased by several percents. However, we show that the effect becomes negligible provided that the lens projected map extends over at least 50 times the Einstein radius. 
   }
   

\maketitle

\section{Introduction}
%\setstcolor{blue}
The diversity and increasing precision of cosmological probes used by the astrophysical community to measure the Hubble constant $H_0$ provides a unique opportunity to test our cosmological paradigm. In past years, a growing tension developed between the $H_0$ inference from methods based on the cosmic microwave background and baryonic oscillations and most of the other measurements \citep{PlanckVI,Abbott2018,BAO_philcox2020,Riess_cepheid,HolicowXIII}. While the multiplicity of probes suggests that systematic errors are not a plausible explanation for the  tension, each technique should continue to be scrutinised for potential systematic errors independently of its inferred value of $H_0$. The time-delay cosmology technique, which uses multiply-imaged strongly lensed quasars to measure $H_0$, offers a poweful means of obtaining the $H_0$ measurement independent of the local distance ladder \citep{HolicowXIII,TDCosmoIV}. The accuracy of this technique relies on our understanding of galaxies and on assumption(s) on their total mass density profiles \citep[e.g.][]{SS2013,RXJ1131,TDCOSMOI}. An important way to validate the technique consists in using simulated lensed systems with various levels of complexity. 
The Time Delay Lens Modelling Challenge has recently been used to test the accuracy of the $H_0$ measurement using mock images of strongly lensed quasars with Hubble Space Telescope (HST) image quality \citep{TDLMCI,TDLMCII}. Mock gravitationally lensed systems based on 'numerical' galaxies have been used for various other applications, including the validation of analysis frameworks, the training of lens-finding algorithms, and studies of galaxy and dark matter properties \citep[e.g.][]{Xu2015,Xu2016, Despali2018, SeagleI, SeagleII, Metcalf2019, Denzel2020, Enzi2020}.

The most commonly used technique to create a mock lens system from simulated galaxies is to extract a mass map from a particle-based simulation and use it to calculate lensing quantities (i.e. lensing potential and its first and second derivatives) needed to emulate the gravitationally lensed images. For this purpose, galaxies from high resolution hydrodynamical simulations, including, for example, EAGLE \citep[Evolution and Assembly of GaLaxies and their Environments,][]{Schaye2015, C15} or Illustris \citep{Illustris,IllustrisDandan}, have been widely used. Different types of software, such as \lenstro \citep{lenstro2018} and GLAMER \citep[Gravitational Lensing Simulations with Adaptive Mesh Refinement,][]{GlamerI}, can handle the inference of lensing quantities from mass maps using fast Fourier transform convolution.
Fast Fourier is a commonly used technique to speed up the calculation of lensing quantities, which imply computationally expansive numerical integration but it remains a demanding procedure \citep{GlamerI, Plazas2020}. One could wonder what mass map resolution should be used and what size of map is relevant to be sufficiently precise in the mock creation while minimising the computational time. One generally considers that a strongly lensed system is determined by the projected mass inner to the lensed images. This would suggest that a region extending over a few Einstein radii ($\theta_E$) is sufficient for the simulations. However, this consists in effectively ignoring any source of shear, and/or perturbations caused by substructures and/or anisotropy in the mass distributions. 
Moreover, depending on the symmetry of the problem, cutting the mass distribution at a given radius not only automatically removes the mass beyond that radius, but it may also introduce numerical artefacts that could wrongly be attributed to properties of the examined lens mass distribution. In this paper we focus on this latter point and quantify the impact of the shape (and size) of the integration domain on the lensing quantities inference.

We used a smooth analytical cored power-law model, including the isothermal case, with conformal isodensity contours, that is, isocontours all having their axis aligned with each other, and discretised the analytical model on a grid to emulate mass maps from numerical simulations. This density profile was chosen to mimic galaxy mass profiles from numerical simulations \citep[e.g.][]{SeagleII,Du2020}. We then truncated those maps and, applying masks spanning a broad range of shapes and extent, used a standard pipeline to create mock lens systems and subsequently modelled them. We find that an 'artificial' shear is created during the mock contruction if the truncation does not follow isodensity contours of the input conformal power-law lens distribution.

In Section \ref{mock_creation}, we explain how a strongly lensed system can be built from a convergence map (i.e. surface mass density  normalised by critical density). In Section \ref{method}, we explain our methodology and synthesise the results in Section \ref{results}. We discuss our results and the potential existence of artificial shear in other works in Section~\ref{discu}. We finally summarise and conclude in Section \ref{conclu}.

Calculations presented in this work assume the following cosmological parameters $\Omega_{\Lambda}=0.693$, $\Omega_{\text{m}}=0.307$, and $H_0=67.77 ~\text{km}~\text{s}^{-1}~\text{Mpc}^{-1}$. This cosmology is the same as the one used in the EAGLE simulations \citep{Schaye2015,EAGLE}.

\section{From convergence maps to lensed images}
\label{mock_creation}

Constructing a mock lens system with time-delays requires one to create a mapping between the source and lens plane. In the case of a single-plane lensing, this is performed through the lens equation \citep[see e.g.][]{Refsdal2,Refsdal1}:
\begin{ceqn}
\begin{align}
\boldsymbol{\beta}=\boldsymbol{\theta}-\boldsymbol{\alpha},
\end{align}
\end{ceqn}
where $\boldsymbol{\beta}$ is the unlensed position, $\boldsymbol{\theta}$ is the observed lensed position, and $\boldsymbol{\alpha}$ is the deflection angle. Moreover, the excess arrival time needed to reach the observer (compared to the unperturbed path) for a light source at an unlensed position $\boldsymbol{\beta}$ is
\begin{ceqn}
\begin{align}
t(\boldsymbol{\theta},\boldsymbol{\beta}) = \frac{D_{\Delta t}}{c} \left( \frac{(\boldsymbol{\theta}-\boldsymbol{\beta})
^2}{2} - \psi(\boldsymbol{\theta}) \right),
\end{align}
\end{ceqn}
where $\psi(\boldsymbol{\theta})$ is the lensing potential at position $\boldsymbol{\theta}$. The time-delay distance $D_{\Delta t}$ is defined as 
\begin{ceqn}
\begin{align}
D_{\Delta t} \equiv (1+z_l)\frac{D_l D_s}{D_{ls}},
\end{align}
\end{ceqn}
where $z_l$ is the lens redshift, $D_l$ is the angular distance between the observer and the lens, $D_s$ is the angular distance between the observer and the source, and $D_{ls}$ is the one between the lens and the source. The lensing potential is defined such that its gradient is equal to the deflection, $\boldsymbol{\nabla} \psi = \boldsymbol{\alpha}$, and half its Laplacian is equal to the convergence (i.e. lens surface mass density normalised by the critical density), $\boldsymbol{\nabla}^2 \psi = 2 \kappa$. The source magnification is given by the determinant of the magnification 2D tensor $\boldsymbol{M}$ with
\begin{ceqn}
\begin{align}
\boldsymbol{M}^{-1}=\boldsymbol{1}-\boldsymbol{\nabla}\boldsymbol{\nabla} \psi.
\end{align}
\end{ceqn} 

We see that to generate an artificial lens system, we need to know the lensing potential $\psi$, its first derivatives (the deflection), and its second derivatives. 
However, the gravitational potential is generally not directly accessible from the numerical simulations, but instead one has access to the surface mass density of the lensing galaxy. Once the source and lens redshifts are chosen, a convergence map can be calculated, normalising the surface mass density by the critical surface density: 
\begin{ceqn}
\begin{align}
\kappa \equiv \frac{\Sigma}{\Sigma_{cr}} \text{~~~with~~~} \Sigma_{cr}=\frac{c^2}{4 \pi G}\frac{D_s}{D_l D_{ls}}.
\end{align}
\end{ceqn} 
Fortunately, retrieving the potential and deflections is possible through \citep{Saas-Fee2006}:
\begin{ceqn}
\begin{align}
\boldsymbol{\alpha}(\boldsymbol{\theta}) &= \frac{1}{\pi}\int_{\rm I\!R^2} d^2 \boldsymbol{\theta'} ~\kappa(\boldsymbol{\theta'})~\frac{\boldsymbol{\theta}-\boldsymbol{\theta'}}{\vert \boldsymbol{\theta} - \boldsymbol{\theta'} \vert^2},\label{alpha}\\ 
\psi(\boldsymbol{\theta}) &=\frac{1}{\pi} \int_{\rm I\!R^2}d^2\boldsymbol{\theta'} ~ \kappa(\boldsymbol{\theta'}) ~\text{ln} \vert \boldsymbol{\theta}- \boldsymbol{\theta'} \vert. \label{pot}
\end{align}
\end{ceqn}

Those integrations are generally performed by means of convolution using a fast Fourier transform (FFT): They both are a convolution of the $\kappa$ map with a kernel and $\kappa \otimes kernel = \text{FT}^{-1}(\text{FT}(\kappa) * \text{FT}(kernel))$.

The second derivatives ($\boldsymbol{\nabla}\boldsymbol{\nabla} \psi$) cannot be obtained the same way as the potential and the deflection. This Hessian matrix is thus constructed through a discrete derivation of the deflection $\boldsymbol{\alpha}$.

\section{Practical specifications} 
\label{method}
To quantify the impact of the truncation on the lensing observables, we used a known analytical density profile, and we discretised and truncated it at different radii and for different shapes of the truncation region. This has allowed us to compare $\boldsymbol{\alpha}$ and $\psi$, which were derived numerically (Sect. \ref{mock_creation}), to their expected values derived analytically. Based on the discretised $\kappa$ map and inferred lensing quantities, a mock lensed system can also be generated and modelled. This modelling step allowed us to quantify the impact of introduced artefacts on model parameters (e.g. external shear).

We used an analytical non-singular isothermal ellipsoid \citep[NIE,][]{NIE} mass distribution and discretised its convergence on a grid. The NIE with its major axis oriented along the x-axis has the following  analytical expressions for $\kappa$, $\psi$, and $\boldsymbol{\alpha}$ :
\begin{ceqn}
\begin{align}
\kappa(x,y) &= \frac{b}{2} \left(q^2(s^2+x^2)+y^2 \right)^{-1/2}\\
\psi(x,y) &= x \, \alpha_x+y \, \alpha_y - b s \frac{1}{2} \ln \left((\phi + s)^2+ (1-q^2)x^2 \right)\\
\alpha_x(x,y) &= \frac{\partial \psi(x,y)}{\partial x} = \frac{b}{\sqrt{1-q^2}} \arctan \left(\frac{\sqrt{1-q^2}x}{\phi+s}\right)\\
\alpha_y(x,y) &= \frac{\partial \psi(x,y)}{\partial y} = \frac{b}{\sqrt{1-q^2}} \text{arctanh} \left(\frac{\sqrt{1-q^2}y}{\phi+q^2s}\right),
\end{align}
\end{ceqn}
\begin{comment}
\begin{eqnarray}
\kappa(x,y) &=& b \left(q^2(s^2+x^2)+y^2 \right)^{-1/2}\\
\psi(x,y) &=& x \, \alpha_x+y \, \alpha_y - b s \frac{1}{2} \ln \left((\phi + s)^2+ (1-q^2)x^2 \right)\\
\alpha_x(x,y) &=& \frac{\partial \psi(x,y)}{\partial x} = \frac{b}{\sqrt{1-q^2}} \arctan \left(\frac{\sqrt{1-q^2}x}{\phi+s}\right)\\
\alpha_y(x,y) &=& \frac{\partial \psi(x,y)}{\partial y} = \frac{b}{\sqrt{1-q^2}} \text{arctanh} \left(\frac{\sqrt{1-q^2}y}{\phi+q^2s}\right)
\end{eqnarray}
\end{comment}
where $b$ is the scale  radius of the convergence (it is equal to the Einstein radius if $s=0$), $q$ is the axis ratio, $s$ is the core radius, and $\phi^2=q^2(x^2+s^2)+y^2$. 
To ease the discussion with literature results, we decided to mimic a galaxy with a morphology and redshift similar to a galaxy extracted from an EAGLE hydro-simulation \citep{Schaye2015, EAGLE}. More specifically, we chose the fiducial parameters  $b = 2 \arcsec$, $s = 0.2\arcsec$, $q=0.7522$, and $PA=-22.5\degree$. In addition, we set the lens redshift to $z_l=0.271$.

We used different grid sizes ranging from 16.1$\arcsec \times$16.1$\arcsec$ to 88.5$\arcsec \times$88.5$\arcsec$. To emulate the truncation, we multiplied the $\kappa$ map with a numerical mask. We considered three different shapes: squared (i.e. taking the $\kappa$ map as it is); circular (i.e. multiplying with a circular mask whose diameter has the size of the $\kappa$ map); and elliptical (i.e. multiplying with an elliptical mask whose major-axis has the size of the $\kappa$ map). 
We used a grid with a pixel size of  0.025$\arcsec$. This resolution is half the typical HST sampling for drizzled images, and it corresponds to 0.1 kpc/pixel in the lens plane. This resolution is high enough so as to not impact our tests: Increasing the resolution by a factor of two yields an offset on the image positions by less than 1 milli-arcsecond, and on relative time delays by less than 1 milli-day.

With this set-up, we implemented the numerical integration\footnote{Within \lenstro, the convolution between the $\kappa$ map and the $kernel$ having twice the $\kappa$ map size is performed using \texttt{scipy.signal.fftconvolve}, a standard FFT convolution algorithm with zero-padding to reach sizes that are a power of 2.} discussed in Sect. \ref{mock_creation}. This allowed us to create, within \lenstro, a model internally labeled 'INTERPOL' \footnote{The 'INTERPOL' model works the following way: It takes the maps (potential and its first and second derivatives) and interpolates between the points if needed using a bivariate spline approximation of order 3 over a rectangular mesh as implemented in \texttt{scipy}.}.

In order to create mock lensed images, we needed to chose a source luminosity profile. We decided to emulate a system similar to those used for time-delay cosmography, namely a lensed quasar. More specifically, we considered a source at redshift $z=2.0$ constituted of a point source and a circular Sersic host, and we simulated a cross configuration image without lens light. To avoid introducing artefacts with a complex point spread function (PSF), we chose to convolve the image with a Gaussian PSF with FWHM = 0.15$\arcsec$. The image was constructed on a 161$\times$161 pixels map with a pixel size equal to 0.05$\arcsec$. 

This system (see left panel of Fig.~\ref{res_fit}) may then be modelled in a standard way \citep[e.g.][]{TDCOSMOI}.
To avoid introducing biases due to degeneracies between source and lens parameters \citep[e.g.][]{Unruh2017}, we fixed the image positions and the scale radius of the Sersic source. 
The only varying parameters during the modelling step are the lensing galaxy parameters, the source centre, and the time delay distance.

In addition, during the modelling, we also introduced an analytical shear model, defined as  $\boldsymbol{\gamma}\equiv \gamma_1+ \text{i} \gamma_2$. The associated lensing quantities are:
\begin{ceqn}
\begin{align}
\psi(x,y)&=\frac{1}{2} \left(\gamma_1~ x^2 + 2 ~ \gamma_2~ xy - \gamma_1~ y^2\right)\\
\alpha_x(x,y)&= \gamma_1 ~x + \gamma_2~ y \label{shear_alphax}\\
\alpha_y(x,y)&=  \gamma_2~ x - \gamma_1~ y\\
\boldsymbol{M}^{-1}&= \begin{pmatrix} 1-\gamma_1 & \gamma_2 \\ \gamma_2 & 1+ \gamma_1 \end{pmatrix} \\
\kappa(x,y) &=  0.
\end{align}
\end{ceqn}

\section{Results}
\label{results}

\begin{figure*}
    \centering
    \includegraphics[width=\textwidth]{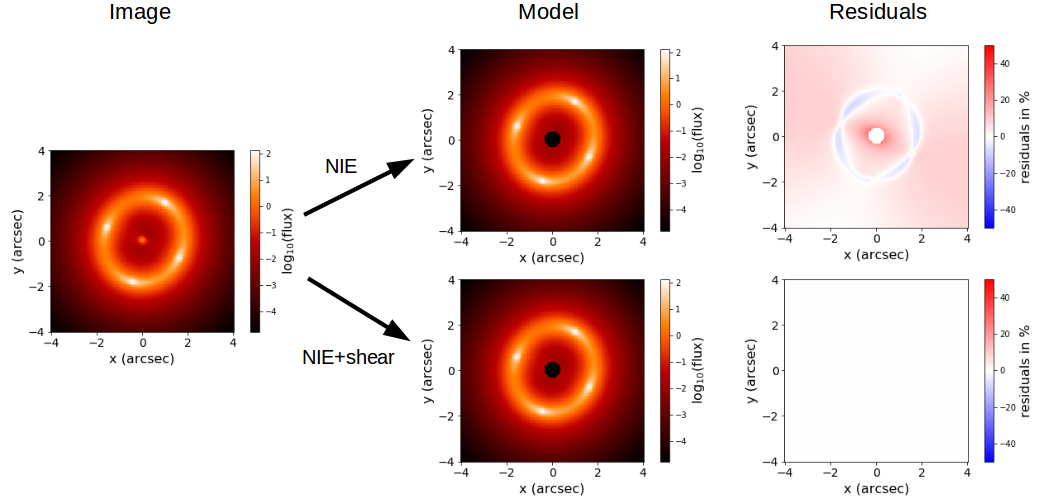}
    \caption{
    Sketch of the method followed to unveil the artificial shear arising due to truncation of the $\kappa$ map: mock lens system (left) created with a pure NIE  $\kappa$ map truncated with a circular mask of diameter 40$\arcsec$ -- i.e. corresponding to a truncation radius of 10\,$\times\,\theta_E$ -- and modelled (middle) using an NIE profile (upper) and NIE+shear (bottom) with associated residuals expressed in the percentage of the input flux (right).}
    
    \label{res_fit}
\end{figure*}

Figure~\ref{res_fit} shows our fiducial mock system (see Sect.~\ref{method}) created with a $\kappa$ map of $40\arcsec$ and a circular mask. One can see that modelling the mock system with our fiducial model leaves substantial residuals (see upper panel of Fig. \ref{res_fit}). In addition, the fitted model parameters, such as the Einstein radius and the ellipticity, are biased compared to the fiducial ones at a level of a few percent. If a shear is fitted in addition to the NIE, we recover the fiducial NIE parameters and a non-zero shear (amplitude of 0.007 in our example). In this case, residuals are compatible with zero (see lower panel of Fig. \ref{res_fit}).
Since this modelled shear does not have any physical meaning (it is an artefact created in the mock process), we dub it 'artificial shear'.

\begin{figure}
    \centering
    \includegraphics[width=0.49\textwidth]{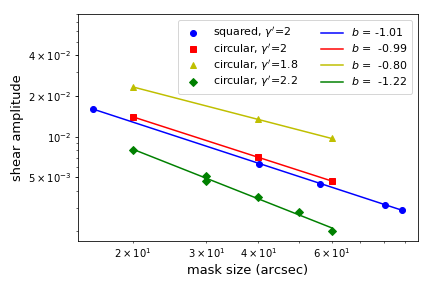}
    \caption{Artificial shear amplitude as a function of the mask size for a given ellipticity module $|e| = 0.14 $. The different markers are for different truncation shapes (circular or squared) and for different slopes of the input cored power-law model $\gamma^\prime$. The solid lines represent the logarithmic linear fits ($\log(y)=a+b\log(x)$). The best-fit values of the slope $b$ are written in the legend.}
    \label{all_cir_sq_slope}
\end{figure}

By modifying the characteristics of our mock lensed systems, we find that the amplitude of this artificial shear depends on (i) the ellipticity of the lens; (ii) the slope of the density profile; (iii) the size of the masking region; and (iv) the shape of the mask. To characterise this external shear, we varied those parameters in a systematic way.  More specifically, we (i) compared the axis ratio of 0.7522 and 0.5590 (i.e. doubling the complex ellipticity module defined as $\vert e \vert = \frac{1-q}{1+q}$, where $q$ is the minor/major axis ratio); (ii) varied the slope of the density profile; (iii) varied the size between 16.1$\arcsec$ and 88.5$\arcsec$; and (iv) compared the shear dependence on the truncation size for circular and squared masks. 
The results of these systematic tests are summarised below:
\begin{itemize}
    \item[i] The artificial shear has an amplitude proportional to the ellipticity module of the input NIE: Doubling the ellipticity module doubles the shear.
    \item[ii] The amplitude of the shear is found to decrease when the density profile is steeper (see Fig. \ref{all_cir_sq_slope}).
    \item[iii] The bigger the mask is, the lower the shear is. We empirically find that $\log(\gamma_{\text{art}})=a+b\log(size)$, where $\gamma_{\text{art}}$ is the artificial shear amplitude and $size$ is the mask size (see Fig. \ref{all_cir_sq_slope}). A least square regression through this relation for each power-law slope $\gamma^\prime$ shows that $b \approx (1-\gamma^\prime)$, in other words, the artificial shear amplitude decreases as $size^{-(\gamma^\prime-1)}$. 
    \item[iv] The dependence of the shear with the size is the same for the square and circular mask (see Fig. \ref{all_cir_sq_slope}).

\end{itemize}
We also note the following properties of the artificial shear:
\begin{itemize}
    \item The artificial shear combines vectorialy with a true external shear: $\boldsymbol{\gamma}_{\text{tot}} =\boldsymbol{\gamma}_{\text{art}} + \boldsymbol{\gamma}_{\text{real}}$.
    \item Using the mock time-delays as an  additional constraint in the likelihood used in lens modelling does not modify the fitting results.
\end{itemize}

In all the cases shown, not accounting for the shear in the modelling yields a biased $H_0$ value. For our fiducial NIE with a mask size of 40$\arcsec$, that is, truncation at 10\,$\times\,\theta_E$, the bias is typically 4-5\% depending on the mask shape. This bias can decrease to 2\% with a mask of 80$\arcsec$. There is, however, no bias when the shear is included in the model. 

Finally, if an elliptical mask, which has an ellipticity module and position angle identical to one of the underlying NIE, is used instead of a squared or circular truncation, no artificial shear is produced. This is true even for small mask sizes (e.g. size = 10$\arcsec$). 

\section{Discussion}
\label{discu}

The previous section demonstrates that to first order, the lensed images are sheared when the truncated $\kappa$ region considered to calculate $\boldsymbol{\alpha}$ and $\psi$ is small and does not follow an isodensity contour. Specifically, a quadrupole (or higher order) contribution, which arises from mass beyond the last complete isodensity contour but is enclosed within the truncation area, is artificially added in the calculation of $\boldsymbol{\alpha}$ and $\psi$, hence producing an artificial shear. 

\begin{figure*}[hbtp]
\centering
\includegraphics[width=\textwidth]{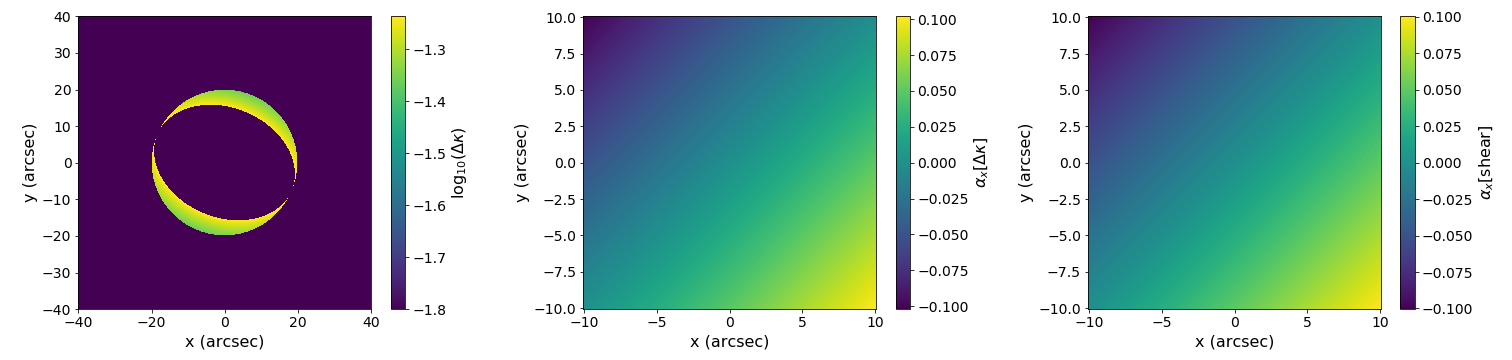}
\caption[]{Comparison of the excess deflection appearing for a circular truncation with the deflection associated to a shear.
Left: $\Delta \kappa$ map for a circular truncation and underlying NIE density profile.
Middle: $\alpha_x$ map (first component of the deflection $\boldsymbol{\alpha}$ vector) created using FFT convolution for eq.~\eqref{alpha} using $\Delta \kappa$. Right: $\alpha_x$ map for a pure shear (eq. \eqref{shear_alphax}). 
}
\label{alpha_from_moon_shear}
\end{figure*}

\begin{figure*}[hbtp]
\centering
\includegraphics[width=\textwidth]{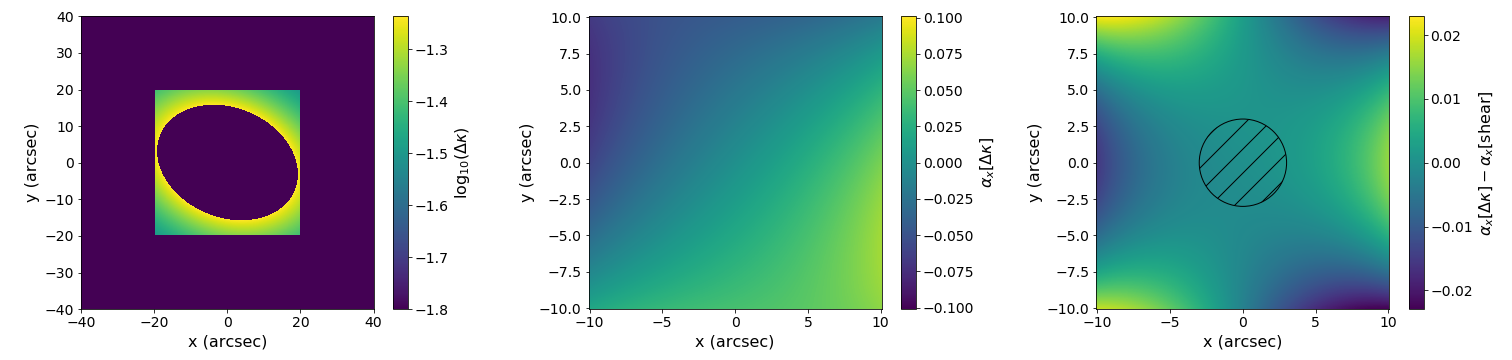}
\caption[]{Comparison of the excess deflection appearing for a squared truncation with deflection associated to a shear. 
Left: $\Delta \kappa$ map for a squared truncation and underlying NIE density profile. 
Middle: $\alpha_x$ map created using FFT convolution for eq. \eqref{alpha} using $\Delta \kappa$. 
Right: difference between $\alpha_x$ map created using $\Delta \kappa$ (middle) and $\alpha_x$ map from a shear ; the hatched region is the $3\arcsec$-radius region where the lensed images are formed.}
\label{alpha_from_squared_shape}
\end{figure*}

We have shown numerically that this 'artificial shear' is effectively a shear term
by calculating the lensing quantities associated to a convergence map $\Delta \kappa = \kappa_{mask} - \kappa_{iso}$, that is to say resulting from the difference between a truncated NIE and 
the same NIE truncated following an iso-density contour.
Fig. \ref{alpha_from_moon_shear} shows that the deflection created with such a $\Delta \kappa$ map is identical to the one produced by a shear. In the case of the squared mask, the deflection created by $\Delta \kappa$ is only approximately the one of a shear in the region where lensed images are located (see Fig.~\ref{alpha_from_squared_shape}) due to the contribution of higher order terms to the deflection.

We have seen that artificial shear could remain unnoticed in simulated lensed systems as it is mostly identical to a real shear.
We identify and discuss hereafter two applications of gravitational lensing simulations where artificial shear has a spurious impact if it is not properly anticipated. 

\subsection{Mocks from simulation to test $H_0$ inference}

When mock lensed systems are created from simulations for the $H_0$ inference, the artificial shear modifies both the image position and the lensing potential, potentially biasing the $H_0$ inference if it is not accounted for. If shear is not fitted when modelling a mock system, we have seen that a bias as large as 2\% exists for a $\kappa$ map truncated at 20 $\theta_E$ for our fiducial NIE. 
We have observed that this bias is also accompanied by substantial residuals, but a common interpretation of the latter would often be to blame the simplicity of the mass model. Residuals might also remain acceptable depending on the signal-to-noise ratio (S/N) added to the mock system, on the contrast between the quasar and its host galaxy as well as on the lens modelling strategy (i.e. point-source only versus full image). 
We are aware of a few published works where numerical simulations were used to investigate the robustness of strong lens modelling on the $H_0$ inference.

The time delay lens modelling challenge \citep[TDLMC,][]{TDLMCII} has invited the community to test their lens modelling techniques on an ensemble of mock lensed systems with time delays. In their so called Rung3, they used lensing galaxies from Illustris (\citealt{Illustris}) and zoom-in (\citealt{zoom_simu}) hydrodynamical simulations. Half of the systems that were produced for Rung3 were built using %They build the lensing quantities using
a custom code \citep{Xu2009_customcode} and cross-checked with the mesh-based FFT algorithm of GLAMER software \citep{GlamerI}; each time  matter distribution up to $R_{200}$ was considered (i.e. the radius at which the average density is equal to 200 times the critical density of the Universe). The other half were built using a ray-tracing code developed by Hilbert \citep{Hilbert2007,Hilbert2009} based on convergence maps extending up to 2 virial radii. 
Since the truncation occurs at typically more than $100\,\theta_E$, one can therefore be confident that a negligible amount of artificial shear is included in those mocks, with a potential impact on $H_0$ below the percent level.

\cite{Tagore2018} used EAGLE \citep{Schaye2015} hydrodynamical simulations to measure the $H_0$ bias from strong lensing and galaxy dynamics. They calculated the deflection and potential using eq. \eqref{alpha} \& \eqref{pot} in their discretised form with a convergence map going up to $R_{200}$ (see Appendix A2 of \citealt{Tagore2018}). 
This work should thus not be affected by artificial shear as described in this analysis.

\subsection{Mocks from simulation to infer galaxy properties}

When mock lensed systems from simulation are used to infer galaxy properties, a systematic shear linked to the galaxy properties may arise. 
We identify a few published works where a spurious shear may be present.

The mock lensed systems created as part of the SEAGLE project \citep{SeagleI} may be affected by artificial shear. SEAGLE aims to study galaxy formation through gravitational lensing using the EAGLE simulation \citep{Schaye2015,EAGLE,SeagleI}. As strong lensing galaxies, \cite{SeagleI} used EAGLE galaxies at redshift z=0.271, sampled over a 161$\times$161 pixels grid with a spatial resolution on the sky of 0.05$\arcsec$ per pixel, that is, a squared truncation of 8$\arcsec\times 8 \arcsec$. 
With such a set-up, we estimate the amplitude of the artificial shear to be $\gamma_{\rm {art}} \sim 0.23 \times |e|$ in the case of an isothermal mass distribution. 
In SEAGLE--I, \cite{SeagleI} find that their lens models require a shear that is proportional to 0.226 $\times \vert e \vert$. They attribute this relation to a degeneracy between shear and ellipticity in lens mass modelling. 
We rather suggest that most of the effect they unveil is in fact caused by an artificial shear. We note that their modelled density profiles should be correctly retrieved and not affected by this issue since shear is included in their mass modelling.

In another work, \cite{Denzel2020} aimed to find ensembles of free-form mass distributions reproducing SEAGLE-simulated lens data and to compare each ensemble to the true input mass profile. The SEAGLE mocks suffer from artificial shear but a shear is allowed in \cite{Denzel2020} modelling. Thus, the retrieved shear should be biased, while the other parameters characterising the mass distribution should be correctly retrieved.

\section{Conclusion}
\label{conclu}

Mass profiles from hydro-dynamical simulations can be used to emulate realistic gravitationally lensed systems. This generally requires deriving the lensing potential, the deflection, and magnification through the integration of the numerical mass distribution. Depending on the size and resolution of the numerical mass density profile, this process can become time-consuming, so that one may consider truncating the mass distribution to a region which encloses the lensed images, that is to say soon beyond the Einstein radius. 
However this truncation, which has been routinely carried out in a circular aperture, would generally introduce numerical artefacts, such as artificial shear, that may be wrongly attributed to intrinsic properties of the mass distribution.

To quantify the biases introduced by truncation, we used elliptical cored power-law models with conformal isodensity contours. 
When the truncation does not follow an isodensity contour, for example when a circular aperture truncation is used, an artificial shear can be created. 
Its amplitude depends on the truncation size as well as on the slope and ellipticity of the lens mass density profile. \cite{SeagleI}, having used a squared truncation of 8$\arcsec \times 8 \arcsec$, that is, $\theta_{\rm{trunc}} = $ 3-4\,$\times\,\theta_E$% for lens systems with $\theta_E \approx 1\arcsec-2 \arcsec$
, report on the existence of a correlation between external shear and ellipticity (see eq. (8) therein) when modelling lensed systems from EAGLE hydro-dynamical simulations. While most of the relation they found should be attributed to spurious shear, the rest of their analysis remains unaffected. 

We discourage truncating $\kappa$ maps at low radius, although this procedure may benefit from minimising the computation needs. Our fiducial isothermal mass distribution, which is characterised by an axis ratio $q \sim  0.75$, can serve as a practical guideline regarding the artificial shear introduced by truncation. Based on our fiducial mass distribution, we recommend a minimum truncation radius of $50 \times \theta_E$ which corresponds to a maximal spurious shear of 0.001. While this seems to be conservative, we stress that not accounting for a shear when modelling such a mock may yield a systematic error on the time-delay as large as 1\%. For a galaxy profile that decreases more steeply or is rounder than this fiducial case, truncation can be performed at smaller radii (see Fig.~\ref{all_cir_sq_slope}). Conversely, truncation should take place farther away for shallower density or galaxies with a smaller axis ratio. 

This work has focused on numerical artefacts introduced by the truncation. However, one should not ignore that truncation also means that some of the mass at the outskirts of the lensing galaxy halo gets removed in the lensing calculation. This mass generally contributes to the lensing plane through a (constant) convergence and (internal) shear \citep[e.g.][]{Keeton1997shearelli}. The impact of internal shear on, for example, the $H_0$ inference is beyond the scope of this work and will be presented in a forthcoming paper.

\section{Acknowledgments}

The authors thank Xuheng Ding and Frédéric Courbin for their wise comments and important scientific discussions on this work.

This work makes use of \lenstro v1.3.0. \citep{lenstro2018} and of the following Python packages : \textsc{Python} v3.6.5 \citep{Python1,Python2}, \textsc{Astropy} v3.0.5 \citep{astropy:2013,astropy:2018}, \textsc{Numpy} v1.17.4 \citep{Numpy}, \textsc{Scipy} v1.3.3 \cite{scipy}, \textsc{Matplotlib} v2.2.3 \citep{Matplotlib}.

This project has received funding from the European Research Council (ERC) under the European Union’s Horizon 2020 research and innovation programme (COSMICLENS : grant agreement No 787886).

DDX acknowledges Tsinghua research funding project no. 20197040016 and the Chinese Academy of Sciences (CAS) project no. 114A11KYSB20170054.

\bibliographystyle{aa} 
\bibliography{biblio}

\end{document}